\documentclass[titlepage,12pt,a4paper,reqno]{article}
\usepackage{amsmath,amssymb,amsfonts,graphics}
\usepackage[linktocpage=true,colorlinks,pdftex]{hyperref}
\usepackage{xcolor}
\colorlet{linkequation}{blue}
\usepackage{bm}
\usepackage{doi}
\usepackage{hyperref}
\hypersetup{colorlinks, citecolor=violet, filecolor=black, linkcolor=black, urlcolor=blue}
\newcommand*{\refeq}[1]{%
  \begingroup
    \hypersetup{
      linkcolor=linkequation,
      linkbordercolor=linkequation,
    }%
    \ref{#1}%
  \endgroup
}
\allowdisplaybreaks

\begin{document} 


\begin{titlepage}

\centerline{\Large \bf 
Graviton spectrum in simplified Dark Matter models} 
\medskip
\centerline{\Large \bf with graviton mediators in the de Sitter space}
\vskip 1cm
\centerline{ \bf Ion V. Vancea }
\vskip 0.5cm
\centerline{\sl Grupo de F{\'{\i}}sica Te\'{o}rica e F\'{\i}sica Matem\'{a}tica}
\centerline{\sl Departamento de F\'{\i}sica}
\centerline{\sl Universidade Federal Rural do Rio de Janeiro}
\centerline{\sl Cx. Postal 23851, BR 465 Km 7, 23890-000 Serop\'{e}dica - RJ,
Brasil}
\centerline{
\texttt{\small ionvancea@ufrrj.br} 
}

\vspace{0.5cm}

\centerline{7 January 2019}

\vskip 1.4cm
\centerline{\large\bf Abstract}

This is the second in a series of papers investigating the formulation of the simplified Dark Matter models with graviton mediators in cosmological backgrounds. We address here the crucial problem of the fundamental observable of interest, namely the graviton spectrum in a FRW cosmological background with an arbitrary Dark Matter background component. We calculate the correction to the free graviton two-point function up to the second order in the coupling constant between the Dark Matter and the graviton in the simplified Dark Matter model with graviton mediators approach in the de Sitter space. Our result is model independent in the sense that it does not depend on the particular form of the Dark Matter fields. Also, due to the universality of the interaction between the Dark Matter and the graviton, the result obtained here apply to the interaction between the baryonic matter and the gravitons. As an application, we discuss in detail the massive scalar Dark Matter model and calculate the first order correction to the two-point function due to two Dark Matter modes in the adiabatic regime.

\vskip 0.7cm 

 Quantum gravity in de Sitter space. 

\noindent

\end{titlepage}


\section{Introduction}

Recently, there has been an increasing number of evidences in favor of the Dark Matter (DM) paradigm from observations of both astrophysical and cosmological nature. These observations include the velocity distribution of the galaxies in clusters and superclusters,  the orbitating bodies inside galaxies  \cite{Zwicky:1937zza} - \cite{Stierwalt:2017qeo}, the estimates in our solar system \cite{Adler:2008ky}, the $X$-ray spectrometry and the luminosity and the gravitational lensing measurements \cite{Ferrari:2008jr}-\cite{Bradac:2008eu}. The basic properties of the DM are consistent with the structure formation, the data from the Lyman-alpha forest, with the computational simulations and the missing satellite problem \cite{Springel:2005} - \cite{Nierenberg:2016nri} (for recent reviews on DM in astrophysics and cosmology see, e. g. \cite{Freese:2008cz,Salucci:2018hqu}). All these observations are in agreement with the  hypothetical non-baryonic DM component of the Universe characterized by $p << \rho$ that exerts a supplementary force on the moving bodies at large scale.
Despite the large amount of data obtained so far, the microscopic nature of the DM is still elusive. The natural working hypothesis is that the DM is composed of non-baryonic particles that interact through the weak interaction, or supersymmetric particles, or axions. The search of these particles at colliders has not produced any new information, yet, about the microscopic structure of the DM mainly due to the vast possibilities to be explored (see for recent models and results, e. g \cite{Arun:2017}-\cite{Boveia:2018yeb} and the references therein). Beside the very general assumptions that the DM should belong to one or more known representations of the Lorentz group, obey the general principles of the Quantum Field Theory and couple weakly with the Standard Model (SM) particles, not much can be added a priori to the search list of the processes involving the DM. Therefore, in order to obtain informations on the quantum structure of the DM, it is crucial to explore thoroughly the landscape of DM interaction models that can be related to the experimental observables.

An important class of models that meet this requirement are represented by the \emph{simplified models} proposed in \cite{Alves:2011wf} which have been investigated during the last years from theoretical as well as phenomenological point of view. In these models, the interaction between the DM and the SM particles, respectively, is carried out by mediators of mass comparable with the interaction energy. The Lagrangian functionals of the simplified models are effective Lagrangians, in the sense that they are obtained after integrating out higher energy degrees of freedom. The interactions are described in terms of collider observables such as particle masses and spins, production cross-sections, branching fractions and decay widths. A review of the recent experimental results obtained at LHC for different types of mediators like mixed couplings to quarks, invisible, vector and axial-vector particles, respectively, can be found in \cite{Albert:2017onk} (see also \cite{Balazs:2017hxh,Arcadi:2017kky}).

The scarcity of data obtained so far at collider experiments and the bulk of astrophysical and cosmological observations that indicate that the DM component interacts much stronger through the gravitational channel than by other mediators, has motivated the generalization of the simplified DM models to spin-2 mediators \cite{Lee:2013bua}-\cite{Kraml:2017atm}. 
In these works, the output of the models is formulated in terms of LHC observables. Therefore, the gravitons are defined in the flat space-time. However, as the analysis of the next generation simplified DM models suggests, it could be experimentally feasible to obtain information about the DM from the data obtained from cosmological and astrophysical observations \cite{Bauer:2016gys,Huang:2018xbi}. To this end, one has to develop models of the interaction between the DM and the gravitons in cosmological backgrounds. 
A first step in this direction was taken in a previous work \cite{Vancea:2018bom}, where the general principles and computational ideas have been proposed for simplified DM models with graviton mediators in a general curved background and in particular in the de Sitter space $dS_4$ by generalizing the constructive principles of the simplified models in the Minkowski space-time \cite{Lee:2013bua,Lee:2014caa}. 

Note that a class of models that substitute the interaction between the DM and the gravity called \emph{Dark Matter emulators} have been criticized on the basis of the observation of gravitational wave GW170817 signal from a merger of a binary neutron star by LIGO-Virgo in the NGC 4993 galaxy \cite{Boran:2017rdn}. This observation correlated with the electromagnetic spectrum from radio to gamma frequencies seems to rule out the models characterized by: i) the coupling between the baryonic matter with the metric perturbed by the presence of the DM and ii) the coupling between the gravitational waves with the metric in the absence of the DM, thus questioning the necessity of DM at all. However, the simplified models discuss here do not fall within the DM emulators class since there is an explicit DM component. Moreover, the coupling between the baryonic matter with the metric is done in the same unperturbed background to which the DM as well as the gravitational waves couple, too.

According to the current view, the DM fields are assumed to have the same geometrical properties as the SM fields, e. g. they belong to the  representations of the local Lorentz group, they obey the covariance principle, etc. Therefore, the constructive principles of the DM models are the same as of similar models involving the baryonic matter. As a matter of fact, in the simplified models the interaction term between the DM and the graviton field is given by the usual covariant coupling between the DM energy-momentum tensor with the linear perturbation $h_{\mu \nu}(x)$ of the background metric $g_{\mu \nu}(x)$ \cite{Lee:2013bua,Lee:2014caa,Vancea:2018bom}. Since that represents an universal property of matter in gravitational field, the characterization of the DM is only through the parameters of the model. Therefore, the theoretical results obtained in this way are valid for all types of matter that obey the covariance principle. Also, due to the same generality, this formalism is model independent since it can accomodate any type of DM field. The results obtained previously in this context \cite{Vancea:2018bom} allow one to calculate the two-point and three-point Green's functions for the interaction between the DM and the gravitons in the de Sitter space. However, the computations from  \cite{Vancea:2018bom} were performed in the Euclidean de Sitter space which is isomorphic to the $S^4$ sphere. The analytic continuation from a curved space-time to its Euclidean version obtained by a Wick-like rotation which is not unique due to the various possible choices of the local and global time-like Killing vectors is a matter of discussions in the literature (see, e. g. \cite{Visser:2017atf}). In the present case, the Euclidean formulation of the simplified models suffers from some drawbacks like a less transparent  of definition of the observables in the genuine $dS_4$. Also, the use of the covariant gauge to define the gravitons, while providing a mathematically more consistent framework for the analysis, can interfere in the formulation of the observables due to the unfixed parameters of the gauge. And finally, there are general problems related to the path integral quantization in cosmological backgrounds \cite{Prokopec:2010be}. On the other hand, the Euclidean formalism has a major advantage since the vacuum state is well defined. 

The main goal of the present paper is to define and calculate a concrete observable for the simplified DM models with graviton mediators in de Sitter backgrounds, namely the graviton spectrum. This represents the most important observable in the cosmology and in the case of free fields in the cosmological background it can be derived from the two-point correlation functions and it can be compared against the CMB data. The free graviton spectrum in different gauges in $dS_4$ can be found in several places in the literature. Our task is to determine the effect of the DM background on the graviton viewed as a quantum linear perturbation of the background metric and calculate the modification induced in the graviton spectrum by the DM fields. 

In the present paper, the above mentioned issues of the simplified models are addressed by working in the $dS_4$ and by choosing the \emph{physical gauge} in which all gauge degrees of freedom of the free graviton are fixed \cite{Ford:1977dj}-\cite{Higuchi:2002sc}\footnote{For free gravitons, the equivalence between the two-point functions in the covariant and physical gauges, respectively, was proved in \cite{Faizal:2011iv}.}. As usual, the quantum excitations are defined with respect to the so called \emph{Euclidean vacuum} which in our case is the direct product of the vacua of all fields of the model. Our result is model independent in the sense that the form of the DM component does not need to be specified. Therefore, as in \cite{Vancea:2018bom}, the same reasoning and results can be applied not only to the DM components from all representations of the local Lorentz group but also to the baryonic matter. The difference between the models describing either DM or baryonic matter is made only by the appropriate choice of parameters in each kind of matter.

The paper is organized as follows. In order to make the presentation self-consistent, we have included the review of several known concepts about the formulation of gravitons and of the quantum fields in curved space-time, most of which taken from \cite{Mukhanov:2005sc,Mukhanov:1990me}. In Section 2 we present the simplified DM model with graviton mediators in a general FRW cosmological background. This is a straightforward generalization of the corresponding model from the Minkowski and the de Sitter spaces, respectively. Also, we review here the construction of the classical graviton field in the physical transverse-traceless gauge. In Section 3, we review the quantization of the free physical graviton field in an arbitrary cosmological background. Since the results are well known in the literature, we mainly quote the relevant relations for the purpose of stating the concepts and the notations. In the same section we derive a new result which is the corrections to the graviton field in the de Sitter space for a general DM component. The choice of the de Sitter space, which is the simplest FRW model, is motivated by the fact that it represents a phenomenological phase of the early Universe. Also, it makes the otherwise general discussion of the simplified models concrete. We show that for any simplified DM model in the de Sitter space there are contributions to the free graviton field from the interaction with the quantum DM energy-momentum tensor. These contributions are computable and the results depend on the specific DM component. In Section 4 we use the corrected graviton derived in the Section 3 to calculate the graviton spectrum in the DM background in the de Sitter space. In Section 5 we discuss the massive scalar DM model in detail. We obtain the general form of the first and second order corrections that apply to a large variety of physical situations. Then we show that these corrections can be calculated using the canonical quantization method in the de Sitter space. Specific results can be obtained on case by case basis by choosing the adequate vacuum state and the corresponding mode functions for the DM scalar field. We exemplify the construction by calculating the correlation function between two DM modes in the adiabatic regime. The last section is devoted to conclusions. We adopt throughout this paper the natural units $\hbar = c = 1$.

\section{Simplified DM model with graviton mediators}

In this section present the simplified DM model with graviton mediators in an arbitrary FRW cosmological background
and review the construction of the classical physical graviton field. That represents a direct generalization of the similar model in the particular case of the de Sitter space. For a discussion of the construction of the model in a general curved space-time see \cite{Vancea:2018bom}. In the presentation of the perturbation theory we have followed mainly the references \cite{Mukhanov:2005sc,Mukhanov:1990me}.

The main idea of the simplified DM models with graviton mediators is that the gravitons interact with the DM component by the covariant coupling of the DM energy-momentum tensor with the graviton field
\cite{Lee:2013bua,Lee:2014caa}. Since the gravitons are defined as the quantized linear perturbations of a given background metric, the interaction between the DM and the gravitons is first order in the perturbation. Therefore, the total linearized action of the model in a cosmological background has the following general form
\begin{align}
S[g,h,X,\phi] 
& = S_0 [g,h] + S_0 [g,X] + S_0 [g,\phi]
\nonumber
\\
& + S_{int}[g,h,X] + S_{int}[g,h,\phi] + S_{int}[g,X,\phi]+ S_{int}[g,h,X,\phi] \, .
\label{total-general-action-sDMmgm}
\end{align}
Here, we have denoted by $g$ the gravitational background expressed by the metric $g_{\mu \nu}$ which is a solution of the Einstein's equations, by $h$ the classical linear perturbations $h_{\mu \nu}$ of $g_{\mu \nu}$ which are small $|h_{\mu \nu}| << 1$, 
by $X$ the DM component $X$ and 
by $\phi$ other matter fields that can be introduced in the model like e. g. the inflaton. We adopt the working hypothesis that the strength of the interaction between the DM component and the matter fields is negligible in comparison with the interaction between the DM fields and the gravitation and thus $ S_{int}[g,X,\phi] = S_{int}[g,h,X,\phi] = 0$. The dynamics of all these fields in the fixed gravitational background is obtained from the variational principle applied to the corresponding free field actions denoted by $S_0$. This dynamics is modified by the mutual interactions described by the actions $S_{int}$. The first order interaction between the DM field and the graviton field is given by the following functional \cite{Lee:2013bua,Lee:2014caa}
\begin{equation}
S_{int}[g,h,X] = -\frac{\alpha}{2} \int d^4 x \sqrt{-g} (x) \,
T^{(X)}_{\mu \nu} (x) \, h_{\mu \nu} (x) \, ,
\label{action-sDMmgm}
\end{equation}
where $\alpha$ is the coupling constant between the DM and the graviton. This is one of the parameters that singles out the DM model from the  otherwise a general covariant interaction action and should be specified in order to describe any particular DM model. Note that by construction, the total action (\refeq{total-general-action-sDMmgm}) should have all the symmetries of its component fields $\Phi (x) = \{g_{\mu \nu}(t), h_{\mu \nu}(x), X(x), \phi(x) \}$  \cite{Vancea:2018bom}. In particular, $S[g,h,X,\phi] $ should be invariant under the gravitational gauge transformations
\begin{equation}
\delta x^{\mu} = \xi^{\mu}(x) \, ,
\qquad
\delta \Phi (x) = - \pounds_{\xi} \Phi (x) \, , 
\label{gravitational-gauge-transformations}
\end{equation}
where $\pounds_{\xi}$ is the Lie derivative with respect to a smooth vector field $\xi (x)$. 

The cosmological backgrounds we are interested in are of the FRW type with the line element of the following form
\begin{equation}
ds^2 = g_{\mu \nu} (t) \, d x^{\mu} d x^{\nu} = dt^2  - a^2 (t) \delta_{i j}
dx^i dx^j = a^2 (\tau ) \eta_{\mu \nu} dx^{\mu} dx^{\nu} \, , 
\label{FRW-metric}
\end{equation}
where $t$ is the co-moving time and $\tau$ is the conformal time, respectively, related to each other by the equation\footnote{
Note that the time variable $t \in \mathbb{R}$ while the conformal time is negative $\tau \in (- \infty , 0 ]$. However, since we need an increasing conformal time for an increasing proper time, we will use whenever necessary $\tau \sim |\tau|$ according to the common practice \cite{Mukhanov:1990me}.}
\begin{equation}
\tau (t) = \int_{t_0}^{t} \frac{d \mathrm{t}}{a(\mathrm{t})}  \, .
\label{co-moving-time}
\end{equation}
The physical processes that took place in the primordial Universe perturbed the gravitational background from the equation (\refeq{FRW-metric}) \cite{Mukhanov:1990me}. Thus, the perturbed line element can be written as
\begin{equation}
ds^2 = \tilde{g}_{\mu \nu} (x) \, d x^{\mu} d x^{\nu}
=\left[ g_{\mu \nu} (t) + \delta g_{\mu \nu} (x) \right] \, d x^{\mu} d x^{\nu}\, .
\label{FRW-metric-perturbed}
\end{equation}
The components of $\delta g_{\mu \nu} (x)$ can be classified according to their transformation properties on spatial Cauchy surfaces as follows: $\delta g_{\mu \mu} (x)$ are scalars, $\delta g_{0 i}(x)$ are vectors and $\delta g_{ij} (x)$ are tensors. 
However, not all the degrees of freedom are physical since the mapping of an arbitrary perturbation $\delta g_{\mu \nu} (x)$ to a different perturbation $\delta {g'}_{\mu \nu} (x)$ by a gravitational gauge transformation from the equation (\refeq{gravitational-gauge-transformations}) is a symmetry of the free graviton equations of motion and should be incorporated to the total action by construction. Indeed, the free action for the graviton field (to be identified with the linear perturbation $\delta g_{\mu \nu} (x)$) is invariant under the reparametrization generated by the scalar fields $\zeta_1$ and $\zeta_2$ given by the relations 
\begin{equation}
t \to t' = \, t + \zeta_{1} \, , \qquad  
x^i \to {x'}^i = x^i + \partial^i \zeta_2 \, .   
\label{reparametrization}
\end{equation}
While the tensor perturbations are invariant under the reparametrization (\refeq{reparametrization}), the scalar components are not.
In particular, one can remove the scalar perturbation $\delta g_{00}(x)$ by using the transformations given by the equations (\refeq{reparametrization}) and fix $\delta g_{00}(x) = \partial_0 \zeta_1$. After that one is left with the following spatial components   
\begin{equation}
\delta g_{ij} (x)  = h_{ij} (x) = 2 \psi (x) \delta_{ij} + 2 \partial_{i} \partial_{j} E(x) + \frac{1}{2}\partial_{(i} E_{j)} (x)  
+ h^{TT}_{ij} (x) \, ,
\label{spatial-perturbation}
\end{equation}
where $\psi (x)$ is a scalar, $F_{i}(x)$ is a transverse vector and $h^{TT}_{ij} (x)$ is a transverse-traceless tensor with respect to the rotation group, that is
\begin{equation}
\partial_{i} F_{i}(x) = 0 \, , 
\qquad
h^{TT}_{ii} (x) = 0 \, , 
\qquad \partial_i h^{TT}_{ij} (x) = 0 \, .
\label{SVT-perturvations}
\end{equation}
One can decompose the gauge parameter $\xi (x)$ into the longitudinal and transversal components
\begin{equation}
\xi_{i} (x) = \partial_{i} \xi (x) + \xi^{T}_{i}(x) \, ,
\qquad
\partial_{i} \xi^{T}_{i} (x) = 0 \, .
\label{xi-decomposed}
\end{equation}
Due to the gauge symmetry from the equation (\refeq{gravitational-gauge-transformations}), the fluctuations $h_{ij} (x)$ are not unique. In order to quantize the system this indeterminancy should be lifted by fixing the gauge at some point during the process of quantization. We adopt the common practice of fixing the gauge already at the classical level which makes it possible to quantize the graviton field by applying canonical methods. The gauge fixing is obtained by imposing the following conditions 
\begin{equation}
\xi (x) = a^2 (t) E(x) \, ,
\qquad
\xi_{0}(x) = - \frac{a(t)}{H} \psi(x) \, ,
\qquad
\xi^{T}_{i} (x) = a^2(t) F_{i} (x) \, .
\label{zero-curvature-gauge}
\end{equation}
If the cosmological model is inflationary it should contain at least one  scalar field $\phi(t)$ whose fluctuations $\delta \phi (x) = \varphi (x)$ should be of the same order of magnitude as $\delta g_{\mu \nu} (x)$. The physical degrees of freedom can be formulated in a gauge invariant manner by introducing the Sasaki-Mukhanov gauge invariant field \cite{Sasaki:1986hm,Mukhanov:1988jd}
\begin{equation}
\bar{\psi} (x) = \psi (x) +  \frac{H}{\dot{\phi}(t)} \phi (x) = 1 - e^{-\mathcal{R}(x)} \, ,
\label{Sasaki-Mukhanov-variable}
\end{equation}
where the dot stands for the derivative with respect to the co-moving time and $\mathcal{R}(x)$ is the curvature perturbation. If the gauge is fixed as in the equation (\refeq{zero-curvature-gauge}), the field $\bar{\psi}(x)$ takes the following form
\begin{equation}
\bar{\psi} (x) = \frac{H}{\dot{\phi}(t)} \phi (x) = \mathcal{R}(x) \, .
\label{zero-curvature-Sasaki-Mukhanov-variable}
\end{equation}
It follows that the physical degrees of freedom of the inflationary model are 
the scalar gravitational potential $\bar{\psi}(x)$ and the tensor graviton field $h^{TT}_{ij} (x)$. The dynamics of the model can be derived from the following action functionals \cite{Mukhanov:2005sc}
\begin{align}
S_0 [g,\bar{\psi}] & = \frac{1}{2} \int d \tau \, d^3 x 
\left( a z \right)^{2}
\left[ 
\left( \bar{\psi}' \right)^2 -
\left( \nabla \bar{\psi} \right)^2
\right] \, ,
\label{action-free-psi}
\\
S_0 [g, h^{TT}] & = \frac{M_{pl}^{2}}{8}\int d \tau \, d^3 x 
\, a^{2}
\left[ 
\left( {h^{TT}_{ij}}' \right)^2 -
\left( \nabla h^{TT}_{ij}  \right)^2
\right] \, ,
\label{action-free-hTT}
\end{align}
where the following notation has been introduced
\begin{equation}
M_{Pl} = \frac{1}{\sqrt{8 \pi G}} \, ,
\qquad
z = \frac{\phi '}{\mathcal{H}} \, , 
\qquad
\mathcal{H} = \frac{a'}{a} \, .
\label{notations}
\end{equation}
Here, $'$ denotes the derivative with respect to the cosmological time and $\mathcal{H}$ is the Hubble's constant in the conformal time. 

The dynamics of the DM component depends on the specific DM model under study. For example, the real scalar DM field has the following free action
\begin{equation}
S_0 [g, X] =  \int d^4 x \sqrt{-g} \,
\left[
\frac{1}{2} g^{\mu \nu} \, \nabla_{\mu} X \, \nabla_{\nu} X 
- V(X) 
\right] \, .
\label{action-DM-scalar}
\end{equation}
The energy-momentum tensor calculated from $S_0 [g, X]$ takes the following form
\begin{equation}
T^{(X)}_{\mu \nu} = \partial_{\mu} X \partial_{\nu} X - 
g_{\mu \nu} 
\left[
\partial^{\rho} \partial_{\rho} X - V(X)
\right] \, .
\label{DM-scalar-energy-momentum}
\end{equation}
Note that the dynamics of the DM field should be defined with respect to the background metric in the action given by the equation (\refeq{action-DM-scalar}). That is in agreement with the simplified model prescription in which the only interaction of the DM field with the background fields and their excitations is through the coupling between the DM energy-momentum tensor taken in the unperturbed background and the graviton field. From that, one can also infer that the field $X(x)$ is unperturbed, otherwise a second interaction between the perturbation of the DM energy-momentum tensor and the background metric would be present in the model.

\section{Gravitons in the presence of DM}

In this section we firstly review the main ideas of the quantization of the graviton field in the physical gauge given by the equation (\refeq{zero-curvature-gauge}) in the inflationary backgrounds following \cite{Mukhanov:1990me}. Then we calculate the correction to the graviton operators in the DM background.

\subsection{Gravitons in inflationary backgrounds}

In the absence of the DM component $X$, the gravitons can be obtained by applying the canonical quantization method to the classical graviton field which is identified with the linear perturbation of the background metric. Since these results are well known, we are just going to quote them here briefly for comparision with the DM background to be analysed latter on. More details can be found in standard texts on cosmology, e. g. \cite{Mukhanov:1990me}.

As shown in the previous section, the graviton degrees of freedom in the inflationary backgrounds are the scalar gravitational potential $\bar{\psi}(x)$ and the tensor graviton field $h^{TT}_{ij} (x)$, respectively. The equation of motion of the field $\bar{\psi}(\tau, \mathbf{x})$ can be obtained by applying the variational principle to the action (\refeq{action-free-psi}) and takes the following form
\begin{equation}
\left[
\frac{\partial^2}{\partial \tau^2} + 
2 \frac{(az)'}{az} 
\frac{\partial}{\partial \tau} - \nabla^2
\right] \bar{\psi} (\tau, \mathbf{x}) = 0 \, .
\label{equation-motion-psi}
\end{equation}
The solution to the equation (\refeq{equation-motion-psi}) can be expanded in Fourier modes because in the conformal time coordinates the spatial leaves of the space-time foliation are mapped conformally into the Euclidean space. The expansion takes the following form
\begin{equation}
\bar{\psi} (\tau, \mathbf{x}) = 
\frac{1}{(2 \pi)^3}\int d^3 \mathbf{k} 
\left[
A(\mathbf{k}) \, \bar{\psi}(\tau, \mathbf{k}) \, e^{i \mathbf{k} \mathbf{x}}
+ 
A^{\dagger}(\mathbf{k})\,  \bar{\psi}^{\star}(\tau, \mathbf{k}) \, e^{-i \mathbf{k} \mathbf{x}}
\right] .
\label{Fourier-expansion-psi}
\end{equation}
The canonically conjugate variable $\bar{\pi} (\tau, \mathbf{x})$ to  $\bar{\psi} (\tau, \mathbf{x})$ derived from the free Lagrangian density is given by the following relation
\begin{equation}
\bar{\pi} (\tau, \mathbf{x}) = \left( a z \right)^2 \bar{\psi}' (\tau,\mathbf{x}) . 
\label{canonical-variable-psi}
\end{equation}
The gravitational potential field is quantized by promoting the Fourier coefficients to operators that act on an Hilbert space and by imposing the following equal-time commutation relations
\begin{equation}
\left[
\bar{\psi}(\tau, \mathbf{x}) , \bar{\pi} (\tau , \mathbf{y} )
\right]
= i \delta (\mathbf{x}-\mathbf{y}) \, .
\label{ETC-psi}
\end{equation}
From the above relation it follows that the mode operators must satisfy the oscillator commutators
\begin{equation}
\left[
A(\mathbf{k}) ,
A^{\dagger}(\mathbf{l})
\right] = (2 \pi)^3 \delta ( \mathbf{k} - \mathbf{l}) 
\, ,
\label{Commutators-A}
\end{equation}
with all other commutators vanishing. The Hilbert space of the quantum gravitational potential is endowed with a Fock space structure in which the states are constructed by applying the creation operators $A^{\dagger}(\mathbf{l})$'s to the vacuum $| \Omega \rangle_{\bar{\psi}}$ that is annihilated by all annihilation operators $A(\mathbf{k})$'s, that is
\begin{equation}
A(\mathbf{k}) | \Omega \rangle_{\bar{\psi}} = 0 \, ,
\label{vacuum-Euclidean-psi}
\end{equation}
for all $\mathbf{k}$. The Fourier analysis of the gravitational potential and the quantization are valid if the function $\bar{\psi}(\tau, \mathbf{k})$ satisfies the following equation of motion in the $\mathbf{k}$-space
\begin{equation}
\left[
\frac{\partial^2}{\partial \tau^2} + \mathbf{k}^2
- \frac{(az)''}{az} 
\right] \left(az \bar{\psi} (\tau, \mathbf{k}) \right)= 0 \, .
\label{equation-motion-psi-k}
\end{equation}
In order to obtain analytical solutions, one needs to make physical assumptions on the potential term from the equation (\refeq{equation-motion-psi-k}). In the slow roll approximation \cite{Mukhanov:1990me}, the equation (\refeq{equation-motion-psi-k}) takes the form of the Bessel's equation and its solution can be expressed in terms of the Hankel functions \cite{Gradshteyn-Ryzhik:2014}
\begin{align}
\bar{\psi} (\tau , \mathbf{k} ) & = \frac{1}{az} \sqrt{\frac{-\pi \tau }{4}} H^{(1)}_{\nu} (-|\mathbf{k}| \tau) \, ,
\label{solution-eq-psi-k-1}
\\
\bar{\psi}^{\star} (\tau , \mathbf{k} ) & = \frac{1}{az} \sqrt{\frac{-\pi \tau }{4}} H^{(2)}_{\nu} (-|\mathbf{k}|\tau) \, ,
\label{solution-eq-psi-k-2}
\end{align}
where 
\begin{equation}
\nu^2 = \frac{(az)''}{az} + \frac{1}{4} \, .
\label{nu-parameter-psi}
\end{equation}
Thus, by establishing the relations (\refeq{ETC-psi}) - (\refeq{nu-parameter-psi}) we have obtained the quantum Sasaki-Mukhanov field which describes the quantization of the curvature perturbation. 

The graviton quantization can be performed in a similar manner. From the action (\refeq{action-free-hTT}), one derives the following equation of motion
\begin{equation}
\left( \Box + 2 \mathcal{H} \right) h^{TT}_{ij} (\tau, \mathbf{x} ) = 0
\, ,
\label{equation-motion-hTT}
\end{equation}
where $\Box $ is the d'Alambert operator in the conformal time in flat space-time. Again, the solution to the equation (\refeq{equation-motion-hTT}) can be expanded in Fourier modes due to the conformal mapping to the Euclidean space and it takes the following form
\begin{equation}
h^{TT}_{ij} (\tau, \mathbf{x} )  = \frac{1}{4 M_{Pl}^2 \pi^3}
\sum_{\sigma = 1, 2} \int d^3 \mathbf{k} \, 
\epsilon^{\sigma}_{ij} (\mathbf{k}) 
\left[
a^{\sigma}(\mathbf{k}) \, \mathrm{h}(\tau, \mathbf{k}) \, 
e^{i \mathbf{k} \mathbf{x}}
+ 
a^{\sigma \dagger}(\mathbf{k})\,  \mathrm{h}^{\star}(\tau, \mathbf{k}) \, e^{-i \mathbf{k} \mathbf{x}}
\right] ,
\label{Fourier-expansion-hTT}
\end{equation}
where $\epsilon^{\sigma}_{ij} (\mathbf{k}) $,  $\sigma =  1, 2 = + , \times $ is the polarization tensor that satisfies the normalization equations
\begin{align}
\sum_{i,j} \, \epsilon^{\sigma}_{ij} (\mathbf{k}) \, \epsilon^{\zeta}_{ij} (\mathbf{k})& = 
\delta^{\sigma \zeta}  \, ,
\label{polarization-normalization-1}
\\
\sum_{\sigma} \epsilon^{\sigma}_{ij} (\mathbf{k}) \, \epsilon^{\sigma}_{kl} (\mathbf{k}) & =
\Pi_{ijkl} (\mathbf{k}) = 
\frac{1}{2} 
\left[ 
P_{ik}(\mathbf{k}) P_{jl}(\mathbf{k}) +
P_{il}(\mathbf{k}) P_{jk}(\mathbf{k}) -
P_{ij}(\mathbf{k}) P_{kl}(\mathbf{k})
\right] \, ,
\label{polarization-normalization-2}
\\
P_{ij}(\mathbf{k}) & = \delta_{ij} - \frac{k_i k_ j}{\mathbf{k}^2} \, .
\label{polarization-normalization-3}
\end{align}
The derivation of the canonically conjugate momentum to the graviton field $\pi^{TT}_{ij}(\tau, \mathbf{x})$ from the action (\refeq{action-free-hTT}) is standard. We obtain the following result
\begin{equation}
\pi^{TT}_{ij} (\tau, \mathbf{x}) = \frac{M^{2}}{4} {h^{TT}_{ij}}' (\tau, \mathbf{x}) \, .
\label{canonical-variable-hTT}
\end{equation}
Due to the above relations, one can immediatly apply the canonical quantization method to the field $h^{TT}_{ij} (\tau, \mathbf{x} )$. As usual, the first step is to promote the fields to operators and to impose the equal-time commutation relations that result from the analysis of constraints \cite{Prokopec:2010be}. Then we obtaine the following relations
\begin{equation}
\left[
h^{TT}_{ij}(\tau, \mathbf{x}) , \pi^{TT}_{kl} (\tau , \mathbf{y} )
\right]
= \frac{i}{2}\left[ 
P_{ik}(\mathbf{k}) P_{jl}(\mathbf{k}) +
P_{il}(\mathbf{k}) P_{jk}(\mathbf{k}) -
P_{ij}(\mathbf{k}) P_{kl}(\mathbf{k})
\right] 
\delta (\mathbf{x}-\mathbf{y}) \, .
\label{ETC-hTT}
\end{equation}
From the above equation it follows that the Fourier coefficients act as creation and annihilation operators of graviton modes on the states of the Fock space of the field $h^{TT}_{ij} (\tau, \mathbf{x} )$ and they satisfy the following commutation relations
\begin{equation}
\left[
a^{\sigma}(\mathbf{k}) ,
a^{\zeta \dagger}(\mathbf{l})
\right] = (2 \pi)^3 \delta^{\sigma \zeta} \delta ( \mathbf{k} - \mathbf{l}) 
\, .
\label{Commutators-B}
\end{equation}
The quantization prescription gives the graviton states that are obtained by acting with the creation operators on the vacuum $| \Omega \rangle_{h}$ defined as the zero eigenvalue solution to the equations
\begin{equation}
a^{\sigma}(\mathbf{k}) | \Omega \rangle_{h} = 0 \, ,
\label{vacuum-Euclidean-h}
\end{equation}
for all $\sigma$ and all $\mathbf{k}$. Like in the case of the curvature perturbations, the solutions to the equation (\refeq{equation-motion-hTT}) and the quantization procedure are consistent if the functions $\mathrm{h}(\tau, \mathbf{k})$ satisfy the following dispersion relation
\begin{equation}
\left[
\frac{\partial^2}{\partial \tau^2} + \mathbf{k}^2
- \frac{a''}{a} 
\right] \left(a \mathrm{h}(\tau, \mathbf{k}) \right)= 0 \, .
\label{equation-motion-hTT-k}
\end{equation}
Finding analytical solutions to the equation (\refeq{equation-motion-hTT-k}) requires more information about the potential term. For example, in the slow roll approximation of the power law inflation with the adiabatic function $\eta (\tau )$, the solutions take the same form as for the scalar gravitational potential and are given by the following relations
\begin{align}
\mathrm{h}(\tau, \mathbf{k}) & = \frac{1}{az} \sqrt{\frac{-\pi \tau }{4}} H^{(1)}_{\nu} (-|\mathbf{k}| \tau) \, ,
\label{solution-eq-hTT-k-1}
\\
\mathrm{h}(\tau, \mathbf{k}) & = \frac{1}{az} \sqrt{\frac{-\pi \tau }{4}} H^{(2)}_{\nu} (-|\mathbf{k}| \tau) \, ,
\label{solution-eq-hTT-k-2}
\end{align}
where $\nu \simeq (3 - \eta)/2$. Thus, we have obtained the explicit form of the graviton operators in the physical gauge and we can consistently construct the observables and the physical states on which these operators act on.

\subsection{Gravitons in DM background in de Sitter space}

In the previous subsection we have reviewed the quantization of the graviton field in a cosmological background that contains a cosmological unperturbed metric and a scalar inflaton. Let us consider now a general DM background $X$ and determine how it affects the graviton operators. Also, in order to make the relations more concrete, we focus on the piece containing the graviton and the DM from the general action (\refeq{action-sDMmgm}). Then it is easy to see that the equation of motion for the spatial transverse traceless graviton in the presence of the DM background takes the following form
\begin{equation}
\frac{M^{2}_{Pl}}{2} 
\left[
a^2 (\tau) 
\Box 
+
2 a (\tau) a' (\tau ) \partial_{\tau}
\right]
h^{TT}_{ij}(\tau, \mathbf{x})
+ \alpha a^4 (\tau) T^{(X)}_{i j} 
(\tau, \mathbf{x})
= 0 \, .
\label{equation-motion-hTT-DM}
\end{equation}
Since the equation (\refeq{equation-motion-hTT-DM}) is in the conformal-time coordinate, the spatial variables belong to a space conformally equivalent to the Euclidean space. Therefore, one can expand the solution to the linear equation (\refeq{equation-motion-hTT-DM}) as well as the DM energy-momentum tensor in terms of Fourier modes as before and write down the following equation
\begin{align}
h^{TT}_{ij} (\tau, \mathbf{x} )  & = 
\int \frac{d^3 \mathbf{k}}{(2 \pi)^3}
\sum_{\sigma = 1, 2}  \, 
\mathrm{F}_{ij}(\tau, \mathbf{k}) \, 
e^{i \mathbf{k} \mathbf{x}} 
\, ,
\label{Fourier-expansion-hTT-DM}
\\
T^{(X)}_{ij} (\tau, \mathbf{x} )  & = 
\int \frac{d^3 \mathbf{k}}{(2 \pi)^3} \,
\mathrm{T}^{(X)}_{ij}(\tau, \mathbf{k}) \, 
e^{i \mathbf{k} \mathbf{x}} 
\, .
\label{Fourier-expansion-DM-energy-momentum}
\end{align}
By substituting the relations (\refeq{Fourier-expansion-hTT-DM}) and (\refeq{Fourier-expansion-DM-energy-momentum}) into the equation of motion (\refeq{equation-motion-hTT-DM}) we obtain the following equation for the coefficients $\mathrm{F}_{ij}(\tau, \mathbf{k})$
\begin{equation}
\left[
\partial^{2}_{\tau} 
+
2 \frac{a' (\tau )}{a (\tau) } \partial_{\tau}
+
\mathbf{k}^2
\right]
\mathrm{F}_{ij}(\tau, \mathbf{k}) 
= - 
\frac{4 \alpha}{M^{2}_{Pl}} a^2 (\tau) \mathrm{T}^{(X)}_{ij}(\tau, \mathbf{k}) \, .
\label{equation-motion-hTT-DM-k}
\end{equation}
In order to obtain more information about the above equation, one has to define a background. Let us specialize on the de Sitter space in what follows, since it represents a phenomenological phase of the early Universe. Then the equation (\refeq{equation-motion-hTT-DM-k}) takes the following form
\begin{equation}
\left[
\partial^{2}_{\tau} 
-
\frac{2}{\tau } \partial_{\tau}
+
\mathbf{k}^2
\right]
\mathrm{F}_{ij}(\tau, \mathbf{k}) 
= - 
\frac{4 \alpha}{M^{2}_{Pl} H^{2}_{0} \tau^2} 
\mathrm{T}^{(X)}_{ij}(\tau, \mathbf{k}) \, ,
\label{equation-motion-hTT-DM-k-1}
\end{equation}
where $H_{0}$ is the Hubble's constant. The equation (\refeq{equation-motion-hTT-DM-k-1}) can be solved by elementary methods (see e. g. \cite{Mikhailov:1978}) and its general solution can be written in terms of Bessel's functions as follows
\begin{align}
\mathrm{F}_{ij}(\tau, \mathbf{k}) & = 
\tau^{\frac{3}{2}} \, 
\left[
C_{1} (\mathbf{k}) J_{\nu} (k \tau ) 
+
C_{2} (\mathbf{k}) Y_{\nu} (k \tau ) 
\right] \epsilon_{ij} (\mathbf{k})
\nonumber
\\
&+ \frac{2 \pi \alpha}{M^{2}_{Pl} H^{2}_{0}} \,
\tau^{\frac{3}{2}} 
\left[
J_{\nu} (k \tau )  
\int d \tau \, \tau^{\frac{1}{2}} Y_{\frac{3}{2}} (k \tau )  
\mathrm{T}^{(X)}_{ij}(\tau, \mathbf{k})
-
Y_{\nu} (k \tau )  
\int d \tau \, \tau^{\frac{1}{2}} J_{\frac{3}{2}} (k \tau )  
\mathrm{T}^{(X)}_{ij}(\tau, \mathbf{k})
\right] \, ,
\label{hTT-DM-F-function}
\end{align}
where $\nu = 3/2$. By substituting the result obtained in the equation (\refeq{hTT-DM-F-function}) into the Fourier expansion of the graviton field from the equation (\refeq{Fourier-expansion-hTT-DM}) and after a short algebra, the following expression is obtained
\begin{align}
h^{TT}_{ij} (\tau, \mathbf{x} )  & = 
\frac{1}{2}
\int \frac{d^3 \mathbf{k}}{(2 \pi)^3}
\sum_{\sigma = 1, 2}  \, 
\tau^{\frac{3}{2}} \,
\left\{
a^{(\sigma)} (\mathbf{k}) H^{(1)}_{\frac{3}{2}} (k \tau ) 
e^{i \mathbf{k} \mathbf{x}} 
+
a^{(\sigma) \dagger} (\mathbf{k}) H^{(2)}_{\frac{3}{2}} (k \tau ) 
e^{-i \mathbf{k} \mathbf{x}}
\right\}
\epsilon^{(\sigma)}_{ij} (\mathbf{k})
\, ,
\nonumber
\\
&+ \frac{2 \pi \alpha}{M^{2}_{Pl} H^{2}_{0}} \,
\tau^{\frac{3}{2}} 
\int \frac{d^3 \mathbf{k}}{(2 \pi)^3}
\left\{
H^{(1)}_{\frac{3}{2}} (k \tau ) 
\left[ 
I^{(1)}_{(\frac{3}{2}) \, ij} (\mathbf{k}; X)
+ i I^{(2)}_{(\frac{3}{2}) \,ij} ( \mathbf{k}; X)
\right]
\right\} e^{i \mathbf{k} \mathbf{x}}
\nonumber
\\
&+ \frac{2 \pi \alpha}{M^{2}_{Pl} H^{2}_{0}} \,
\tau^{\frac{3}{2}} 
\int \frac{d^3 \mathbf{k}}{(2 \pi)^3}
\left\{
H^{(2)}_{\frac{3}{2}} (k \tau ) 
\left[ 
I^{(1)}_{(\frac{3}{2}) \, ij} (\mathbf{k}; X)
- i I^{(2)}_{(\frac{3}{2}) \,ij} ( \mathbf{k}; X)
\right]
\right\} e^{i \mathbf{k} \mathbf{x}}
\, ,
\label{Fourier-expansion-hTT-DM-final}
\end{align}
where we have introduced the following shorthand notation for the integrated operators
\begin{align}
I^{(1)}_{(\frac{3}{2}) \, ij} (\mathbf{k}; X) & = 
\int d \tau \, \tau^{\frac{1}{2}} Y_{\frac{3}{2}} (k \tau ) \,  
\mathrm{T}^{(X)}_{ij}(\tau, \mathbf{k})
 \, ,
\label{integral-I-1-DM}
\\
I^{(2)}_{(\frac{3}{2}) \, ij} (\mathbf{k}; X) & = 
\int d \tau \, \tau^{\frac{1}{2}} J_{\frac{3}{2}} (k \tau ) \,  
\mathrm{T}^{(X)}_{ij}(\tau, \mathbf{k})
 \, .
\label{integral-I-2-DM}
\end{align}
We recognize in the first line of the equation (\refeq{Fourier-expansion-hTT-DM-final}) the free graviton field in the de Sitter space expanded in terms of oscillating modes \cite{Prokopec:2010be}. The next two lines are contributions to this field from the DM sector through the simplified model interaction. 

The relations (\refeq{Fourier-expansion-hTT-DM-final}) - (\refeq{integral-I-2-DM}) show that one can quantize the field $h^{TT}_{ij} (\tau, \mathbf{x} )$ as was done in the previous subsection to obtain the free gravitons and then add the contributions from the quantum DM component $X$ to obtain the correction from the DM background. Thus, we have obtained general instructions on how to calculate the properties of the graviton in the de Sitter space for any DM background. The concrete expressions depends of the particular choice of the DM model which should be viewed as a quantum field in the de Sitter space.

\section{Graviton spectrum in DM background in de Sitter space}

The results obtained in the previous section allow one to define and calculate observables from the quantum graviton fields in the presence of the DM fields in the de Sitter space. The most important cosmological observable is the graviton spectrum which measures the fluctuations of the fields in the cosmological background. Let us write for simplicity the equation (\refeq{Fourier-expansion-hTT-DM-final}) in the following obvious notation
\begin{equation}
h^{TT}_{ij} (\tau, \mathbf{x} )  = 
h^{TT}_{ij, 0} (\tau, \mathbf{x} ) + 
\frac{2 \pi \alpha}{M^{2}_{Pl} H^{2}_{0}} \tau^{\frac{3}{2}} I_{ij} (\tau , \mathbf{x} ; X) .
\label{hTT-simplified}
\end{equation}
Then we can immediately write the two-point correlation function for the graviton as follows
\begin{align}
& \langle \Omega | h^{TT}_{ij} (\tau, \mathbf{x} ) h^{TT}_{mn} (\tau, \mathbf{y} ) | \Omega \rangle  =
\langle \Omega | h^{TT}_{ij, 0} (\tau, \mathbf{x} ) h^{TT}_{mn, 0} (\tau, \mathbf{y} ) | \Omega \rangle
\nonumber
\\
& + \frac{2 \pi \alpha}{M^{2}_{Pl} H^{2}_{0}} \tau^{\frac{3}{2}}
\left[
\langle \Omega | I_{ij} (\tau , \mathbf{x} ; X) h^{TT}_{mn, 0} (\tau, \mathbf{y} ) | \Omega \rangle
+
\langle \Omega | h^{TT}_{ij, 0} (\tau, \mathbf{x} ) I_{mn} (\tau , \mathbf{y} ; X) | \Omega \rangle
\right]
\nonumber
\\
& +
\frac{4 \pi^2 \alpha^2}{M^{4}_{Pl} H^{4}_{0}} \tau^{3}
\langle \Omega | I_{ij} (\tau , \mathbf{x}; X) I_{mn} (\tau , \mathbf{y}; X) | \Omega \rangle 
\, ,
\label{Graviton-Spectrum}
\end{align}
where
\begin{equation}
| \Omega \rangle =  | \Omega \rangle_{h}
\, | \Omega \rangle_{X} \, ,
\label{vacuum-total}
\end{equation}
is the total vacuum of the system. The first line provides the free graviton spectrum $\mathcal{P}_{h} (k, \tau)$ which takes the following form in the de Sitter space
\begin{equation}
\int d k \, \mathcal{P}_{h} (k, \tau) \, 
\frac{\sin (k, |\mathbf{x}|)}{k^2 |\mathbf{x}|} \Pi_{ijkl} (\mathbf{k})
\, ,
\qquad
\mathcal{P}_{h} (k, \tau) = \frac{4 k^3}{M^{2}_{Pl} \pi^2} 
|\mathrm{h}(\tau, \mathbf{k}) |^2 .
\label{spectrum-free-hTT}
\end{equation}
The second line gives the correction to the two-point function at first order and the second line gives the correction at the second order
in the coupling constant $\alpha$, respectively. We note that the second order corrections are pure DM terms, with no contribution from the gravitons. As before, one should substitute the energy-momentum tensor of a specific DM model to obtain concrete relations.

\section{The case of the scalar DM field}

In order to exemplify the above model, let us consider the massive real scalar DM field. The general action given by the equation (\refeq{action-DM-scalar}) can be written in the conformal time and it takes the following form
\begin{equation}
S_0 [X] = \frac{1}{2} \int d \tau d^3 x 
\left[
(X')^2 - (\nabla X)^2 - m^{2}_{eff} (\tau ) X^2
\right] \, ,
\label{DM-scalar-action-conformal-time}
\end{equation}
where we have chosen a concrete form for the potential term that corresponds to a massive DM field\footnote{For an early discussion of the quantum fields on the de Sitter space and of the mass definition see \cite{Grensing:1977fn}.}. The effective mass in a FRW cosmological background has the following form
\begin{equation}
m^{2}_{eff} (\tau ) = m^2 a^2 (\tau ) - \frac{a''(\tau)}{a(\tau )} \, .
\label{DM-scalar-effective-mass}
\end{equation}
In the de Sitter space it is given by the following relation
\begin{equation}
m^{2}_{eff} (\tau ) = 
\left(
\frac{m^2}{H^{2}_{0}}
- 2
\right) \frac{1}{\tau^2} 
\simeq - \frac{2}{\tau^2}
\, ,
\label{DM-scalar-effective-mass-dS}
\end{equation}
where the last approximation is valid for particles with mass much lower than the Hubble's present time constant. By applying the variational principle to the action (\refeq{DM-scalar-action-conformal-time}) the following equation of motion is obtained
\begin{equation}
X'' - \Delta X + m^{2}_{eff} (\tau ) X = 0 \, .
\label{DM-scalar-eq-motion-conformal-time}
\end{equation}
The energy-momentum tensor can be calculated by using the equation (\refeq{DM-scalar-energy-momentum}). 
However, since the gravitons have been defined in the transverse traceless gauge in which only their spatial components are non-zero, one has to determine only the spatial components of the energy-momentum tensor, too. To this end, recall that the spatial part of the perturbed  metric $g_{ij} = \delta_{ij} - h_{ij}$ does not depend explicitly on the space-like coordinates. Moreover, it is of the pure trace form so that it vanishes under the transverse-traceless condition. 
This implies that the energy-momentum tensor given by the equation (\refeq{DM-scalar-energy-momentum}) simplifies to the following expression
\begin{equation}
T^{(X)}_{ij} = \partial_{i} X \partial_{j} X \, .
\label{DM-scalar-energy-momentum-1}
\end{equation}
The components of the energy-momentum tensor in the momentum space can be determined by applying the Fourier expansion to the field $X$. 
From the equation of motion (\refeq{DM-scalar-eq-motion-conformal-time})
we see that the field $X$ and its space-like derivatives can be expanded in to Fourier series as follows 
\begin{align}
X(\tau, \mathbf{x} ) & = \int \frac{d^3 \mathbf{k}}{(2 \pi)^3} \, 
\mathrm{X}(\tau , \mathbf{k}) \, e^{i \mathbf{k} \mathbf{x}} \, ,
\label{DM-scalar-Fourier-1}
\\
\partial_j X(\tau, \mathbf{x} ) & = i \int \frac{d^3 \mathbf{k}}{(2 \pi)^3} \, 
\mathrm{X} (\tau , \mathbf{k} ) \, k_j \, e^{i \mathbf{k} \mathbf{x}} \, .
\label{DM-scalar-Fourier-2}
\end{align}
Here, it is required that the mode functions satisfy the Sasaki-Mukhanov equation \cite{Mukhanov:1988jd}
\begin{equation}
\partial^{2}_{\tau} \mathrm{X}(\tau , \mathbf{k} ) + \omega^{2}_{\mathbf{k}}
 \mathrm{X} (\tau , \mathbf{k} ) = 0 \, ,
\qquad
\omega^{2}_{\mathbf{k}} = k^2 + m^{2}_{eff} (\tau ) \simeq
 k^2 - \frac{2}{\tau^2} \, .
\label{DM-scalar-SM-eq}
\end{equation}
By using the equations (\refeq{DM-scalar-Fourier-1}) and (\refeq{DM-scalar-Fourier-2}), respectively, in the equation (\refeq{DM-scalar-energy-momentum-1}) we obtain after a short algebra the space-like component of the energy-momentum tensor in the momentum-space
\begin{equation}
\mathrm{T}^{(X)}_{ij} (\tau , \mathbf{k} ) = -  
\int d^3 p \, p_i \left( k_j - p_j \right)
\mathrm{X} (\tau , \mathbf{p}) 
\mathrm{X} (\tau , \mathbf{k} - \mathbf{p}) \, . 
\label{DM-scalar-energy-momentum-final}
\end{equation}
By plugging the equation (\refeq{DM-scalar-energy-momentum-final}) into the equation (\refeq{Graviton-Spectrum}) one obtains the following general form of the first order corrections
\begin{align}
C^{(1)}_{ij, mn} (\tau, \mathbf{x}, \mathbf{y} ; X ) & \sim
\int \frac{d^3 k}{(2 \pi)^3} 
\int \frac{d^3 p}{(2 \pi)^3} 
\int d \tau' \, {\tau '}^{\frac{1}{2}}
\left[
Y_{\frac{3}{2}} (k \tau ) J_{\frac{3}{2}} (k \tau' )
-
J_{\frac{3}{2}} (k \tau ) Y_{\frac{3}{2}} (k \tau' )  
\right]
\nonumber
\\
& \times
p_i \left( k_j - p_j \right)
\langle
\mathrm{X} (\tau' , \mathbf{p}) \,
\mathrm{X} (\tau' , \mathbf{k} - \mathbf{p}) \,
h^{TT}_{mn, 0} (\tau, \mathbf{y} )
\rangle \, e^{i \mathbf{k} \mathbf{x}} \, ,
\label{DM-scalar-correction-1-1}
\\
C^{(2)}_{ij, mn} (\tau, \mathbf{x}, \mathbf{y} ; X ) & \sim
\int \frac{d^3 k}{(2 \pi)^3} 
\int \frac{d^3 p}{(2 \pi)^3} 
\int d \tau' \, {\tau '}^{\frac{1}{2}}
\left[
Y_{\frac{3}{2}} (k \tau ) J_{\frac{3}{2}} (k \tau' )
-
J_{\frac{3}{2}} (k \tau ) Y_{\frac{3}{2}} (k \tau' )  
\right]
\nonumber
\\
& \times
p_m \left( k_n - p_n \right)
\langle
h^{TT}_{ij, 0} (\tau, \mathbf{x} )
\mathrm{X} (\tau' , \mathbf{p}) \,
\mathrm{X} (\tau' , \mathbf{k} - \mathbf{p}) \,
\rangle \, e^{i \mathbf{k} \mathbf{y}} \, ,
\label{DM-scalar-correction-1-2}
\end{align}
where the correlators are defined by the equation (\refeq{Graviton-Spectrum}) and $\sim$ denotes the fact that the constant in front of the correlators has been omitted. The second order correction is given by the following equation
\begin{align}
C^{(3)}_{ij, mn} (\tau, \mathbf{x}, \mathbf{y} ; X ) & \sim
\int \frac{d^3 k^{(1)}}{(2 \pi)^3} 
\int \frac{d^3 k^{(2)}}{(2 \pi)^3} 
\int \frac{d^3 p^{(1)}}{(2 \pi)^3} 
\int \frac{d^3 p^{(2)}}{(2 \pi)^3} 
\int d \tau' \, {\tau '}^{\frac{1}{2}}
\int d \tau'' \, {\tau ''}^{\frac{1}{2}}
\nonumber
\\
& \left[
Y_{\frac{3}{2}} (k \tau ) J_{\frac{3}{2}} (k \tau' )
-
J_{\frac{3}{2}} (k \tau ) Y_{\frac{3}{2}} (k \tau' )  
\right]
\left[
Y_{\frac{3}{2}} (k \tau ) J_{\frac{3}{2}} (k \tau'' )
-
J_{\frac{3}{2}} (k \tau ) Y_{\frac{3}{2}} (k \tau'' )  
\right]
\nonumber
\\
& \times
p^{(1)}_{i} 
p^{(2)}_{m}
\left( k^{(1)}_{j} - p^{(1)}_{j} \right)
\left( k^{(2)}_{n} - p^{(2)}_{n} \right)
e^{i \mathbf{k}^{(1)} \mathbf{x}} 
e^{i \mathbf{k}^{(2)} \mathbf{y}}
\nonumber
\\
& \langle
\mathrm{X} (\tau' , \mathbf{p}^{(1)}) \,
\mathrm{X} (\tau' , \mathbf{k}^{(1)} - \mathbf{p}^{(1)}) \,
\mathrm{X} (\tau'' , \mathbf{p}^{(2)}) \,
\mathrm{X} (\tau'' , \mathbf{k}^{(2)} - \mathbf{p}^{(2)}) \,
\rangle \,  .
\label{DM-scalar-correction-2}
\end{align}
Note that the conformal time argument from the solutions (\refeq{DM-scalar-correction-1-1}), (\refeq{DM-scalar-correction-1-2}) and (\refeq{DM-scalar-correction-2}) is positive $\tau \sim |\tau |$.

In what follows we are going to find the first order correlators. The second-order ones do not depend on the gravitons but they can be derived in a similar manner. One way to calculate the correlators is to use the mode expansion of the fields $X$ and $h^{TT}_{ij,0}$. For the free graviton, the mode expansion can be read off of the first line of the equation (\refeq{Fourier-expansion-hTT-DM-final}). The mode components of the field $X$ have the following form
\begin{equation}
\mathrm{X} (\tau , \mathbf{k}) = 
\frac{1}{\sqrt{2}}
\left[
b^{-}_{\mathbf{k}} \mathcal{X}^{*}_{\mathbf{k}} (\tau ) +
b^{+}_{-\mathbf{k}} \mathcal{X}_{\mathbf{k}} (\tau )
\right] \, .
\label{DM-scalar-mode-X}
\end{equation}
The complex mode functions $\mathcal{X}_{\mathbf{k}} (\tau )$ are linearly independent solutions to the Sasaki-Mukhanov equation (\refeq{DM-scalar-SM-eq}). Upon quantization, the coefficients 
$b^{-}_{\mathbf{k}}$ and $b^{+}_{-\mathbf{k}}$ are promoted to mode operators bounded to satisfy the canonical commutation relations 
\begin{equation}
\left[
b^{-}_{\mathbf{p}} , b^{+}_{\mathbf{k}}
\right] = \delta (\mathbf{p} - \mathbf{k}) \, ,
\label{DM-scalar-commutator}
\end{equation}
the rest of commutators being zero. The vacuum state $| \Omega \rangle_X $ is mapped to zero by all annihilation operators $b^{-}_{\mathbf{p}} \, | \Omega \rangle_X = 0$. The quantized fields can be used to calculate
the correlations given by the equations (\refeq{DM-scalar-correction-1-1}) and (\refeq{DM-scalar-correction-1-2}), respectively which should be plugged into the second line of the equation (\refeq{Graviton-Spectrum}). After a somewhat lengthy but straightforward algebra, we obtain the following first order correction to the graviton two-point correlation function
\begin{align}
{\mathcal{A}}^{(1)}_{ij,mn} (\tau , \mathbf{x},  \mathbf{y} ; X) & =
\frac{\alpha}{(2 \pi)^2 M^{2}_{Pl} H^{2}_{0}} 
\tau^{3}
\int \frac{d^3 k}{(2 \pi)^3} 
\int \frac{d^3 p}{(2 \pi)^3}
\int d \tau' \, {\tau '}^{\frac{1}{2}}
\nonumber
\\
&\times \left[
J_{\frac{3}{2}}(k \tau) Y_{\frac{3}{2}} (k \tau )
J_{\frac{3}{2}} (k \tau' ) \mathcal{X}_{\mathbf{p}} (\tau' )
-
J^{2}_{\frac{3}{2}}(k \tau) 
Y_{\frac{3}{2}} (k \tau' ) 
\right]
\mathcal{X}_{\mathbf{p}} (\tau' )
\mathcal{X}_{\mathbf{k}-\mathbf{p}} (\tau' )
\nonumber
\\
& \times 
\left[
p_i \left( k_j - p_j \right) 
\sum_{\sigma = 1, 2} \epsilon^{\sigma}_{mn} (-\mathbf{k}) 
e^{i \mathbf{k} (\mathbf{x} - \mathbf{y})}
+
p_m \left( k_n - p_n \right) 
\sum_{\sigma = 1, 2} \epsilon^{\sigma}_{ij} (\mathbf{k}) 
e^{-i \mathbf{k} (\mathbf{x} - \mathbf{y})}
\right]
\, .
\label{DM-scalar-correction-1-1-final}
\end{align}
The relation (\refeq{DM-scalar-correction-1-1-final}) is the most general form of ${\mathcal{A}}^{(1)}_{ij,mn} (\tau , \mathbf{x},  \mathbf{y} ; X)$ from which one can calculate the first order correction for different scenarios either in the primordial cosmology or at a latter time. In each of this cases the vacuum state and the corresponding mode functions of the DM field should be chosen accordingly. 

For example, if the interaction between the DM and the graviton field is considered in the cosmological de Sitter scenario, then one should pick up the Bunch-Davies vacuum which in the remote past coincides with the vacuum in the Minkowski space-time. It follows that the mode functions corresponding to the Bunch-Davies vacuum should have the asymptotic behaviour of the mode functions in the Minkowski space-time. The standard mode functions with this property are the following ones
\cite{Mukhanov:1990me}
\begin{equation}
\mathcal{X}_{\mathbf{k}} (\tau ) = 
\sqrt{\frac{\pi \tau}{2}} H^{(2)}_{\nu} (k \tau) \, ,
\label{DM-scalar-mode-function}
\end{equation}
where $H^{(2)}_{\nu}(z)$ is the Hankel function and $\nu = \sqrt{\frac{9}{4} - \frac{m^2}{H^{2}_{0}}}$.

Beside the choice of the vacuum and of the mode functions there are other physical conditions that must be taken into account. For example, if the DM modes $\mathbf{k}$ are smaller than the horizon at the initial conformal time $\tau_i$, that is $k \tau_i >> 1$ then the mode functions (\refeq{DM-scalar-mode-function}) can be approximated by their asymptotic form until the horizon crossing time $\tau_k = k^{-1}$ is reached which is the \emph{adiabatic approximation} \cite{Mukhanov:1990me}
\begin{equation}
\mathcal{X}_{\mathbf{k}} (\tau ) \simeq \frac{1}{k} e^{-ik \tau} \, .
\label{DM-scalar-mode-function-adiabatic-approx}
\end{equation}
Since the adiabatic approximation depends on the modes, it follows that the calculations involved in ${\mathcal{A}}^{(1)}_{ij,mn} (\tau , \mathbf{x},  \mathbf{y} ; X)$ should be done with care. In order to exemplify this point let us consider two modes $\mathbf{k}$ and $\mathbf{p}$ that are in the adiabatic approximation and have very close horizon crossing time. Also, consider the approximation in which the modes are smaller than the horizon along the observed interval $[ \tau_1 , \tau_2 ]$.  
Then  $\mathbf{k}$ and $\mathbf{p}$ contribute to 
${\mathcal{A}}^{(1)}_{ij,mn} (\tau , \mathbf{x},  \mathbf{y} ; X)$ by the following amplitude
\begin{align}
{\mathcal{A}}^{(1)} _{ij,mn}(\mathbf{k},\mathbf{p})
& =
\frac{\alpha \sqrt{2}}{4 \pi^3 \sqrt{\pi} \, M^{2}_{Pl} H^{2}_{0}} 
\frac{\tau^{2}}{ k (|\mathbf{k} - \mathbf{p}| + p - k) 
\sqrt{ p \, k |\mathbf{k} - \mathbf{p}|}}
\nonumber
\\
&\times 
\left[
\left( \cos (2 k \tau) + 1 \right)
\cos \left(|\mathbf{k} - \mathbf{p}| + p - k \right)
-
\sin (2 k \tau)
\sin \left(|\mathbf{k} - \mathbf{p}| + p - k \right)
\right] \vert^{\tau_{2}}_{\tau_{1}}
\nonumber
\\
& \times 
\left[
p_i \left( k_j - p_j \right) 
\sum_{\sigma = 1, 2} \epsilon^{\sigma}_{mn} (-\mathbf{k}) 
e^{i \mathbf{k} (\mathbf{x} - \mathbf{y})}
+
p_m \left( k_n - p_n \right) 
\sum_{\sigma = 1, 2} \epsilon^{\sigma}_{ij} (\mathbf{k}) 
e^{-i \mathbf{k} (\mathbf{x} - \mathbf{y})}
\right]
\, ,
\label{DM-scalar-correction-1-1-final-k-p}
\end{align}
where the integration limits are between $\tau_i \leq \tau_1 < \tau_2 \leq \tau_k \simeq \tau_p $ in order for the adiabatic approximation 
given by the equation (\refeq{DM-scalar-mode-function-adiabatic-approx})
to be valid. Similar considerations can be made in other cases where different limits are taken, but its likely that the computational methods have a greater role in calculating the amplitudes.

\section{Conclusions}

In this paper, we have obtained workable relations for the computation of the graviton spectrum in any arbitrary DM background in the simplified DM model with graviton mediators approach in the de Sitter space. The graviton spectrum represents the most important cosmological observable and it is contained in the two-point function of graviton.
Here, we have obtained its general form in an arbitrary DM background  in the equation (\refeq{Graviton-Spectrum}) in which the interaction between the graviton and the DM field introduces corrections at first and second order in the coupling constant, respectively. Also, these corrections correspond to the first and second powers of square Planck mass and square present time Hubble radius. The result is expressed in terms of integrated operators that should be calculated from the DM energy-momentum operator. Next, we have discussed in detail the case of the massive scalar DM field to which we have applied the general relations to determine the form of the first and second order corrections to the graviton two-point function. By using the known results on the quantization of the scalar field in the de Sitter space-time, we have calculated the first order correction in the equation (\refeq{DM-scalar-correction-1-1-final}). Our result describes a general range of possible physical settings defined by specific choices of the vacuum states and the mode functions for the scalar DM field as well as by different physical constraints. In the case of DM modes that obey the adiabatic approximation, the correlation function was calculated explicitly and the result was given in the equation (\refeq{DM-scalar-correction-1-1-final-k-p}).

The results obtained in this paper represent a generalization of the model introduced in the previous work \cite{Vancea:2018bom}. Also, we have improved here on the formulation of the simplified DM models with graviton mediators in the de Sitter space discussed there and which presents difficulties related to the formulation of the path integral and the definition of the observable phenomenological quantities mainly due to the use of the Euclidean de Sitter space. In the present paper we have given a formulation entirely in the conformal time coordinates of the de Sitter space. It is worth noting that even if the main relations have been presented in the de Sitter space for the concreteness, many important steps have been formulated in arbitrary FRW backgrounds. Indeed, it is possible to work out the two-point function of graviton in other cosmological backgrounds and to generalize the formalism by introducing different background fields. Also, due to the universality of the coupling between the DM and the graviton, our results apply verbatim to the baryonic matter, too. We hope to report on this topics soon. 

\section*{Acknowledgements}
It is a pleasure to acknowledge M. C. Rodriguez for discussions and to the anonymous reviewers for providing very insightful comments and suggestions.


\end{document}